# Absolute frequency measurements for hyperfine structure determination of the R(26) 62-0 transition at 501.7 nm in molecular iodine

Andrei GONCHAROV[1,2], Olivier LOPEZ[1], Anne AMY-KLEIN[1], Frédéric DU BURCK[1]

[1] Laboratoire de Physique des Lasers, UMR 7538 CNRS, Université Paris 13,

99 av. J.-B. Clément, 93430 Villetaneuse, France

[2] Institute of Laser Physics,

Siberian Branch of the Russian Academy of Sciences,

Pr. Lavrentyeva 13/3, 630090, Novosibirsk, Russia

**Abstract**

The absolute frequencies of the hyperfine components of the R(26) 62-0 transition in molecular iodine at 501.7 nm are measured for the first time with an optical clockwork based on a femtosecond laser frequency comb generator. The set-up is composed of an $Ar^+$ laser locked to a hyperfine component of the R(26) 62-0 transition detected in a continuously pumped low-pressure cell (0.33 Pa). The detected resonances show a linewidth of 45 kHz (half-width at half-maximum). The uncertainty of the frequency measurement is estimated to be $\sigma = 250\,\text{Hz}$.

## Introduction

The numerous iodine absorption lines in the 0.5 – 0.9 μm spectral range are widely used for laser stabilization in the realization of optical length standards. In particular, seven radiations recommended by the CIPM for the realization of the metre correspond to iodine transitions [1]. The iodine absorption spectrum is composed of the hyperfine structure of rovibrational transitions between the ground electronic state X ($^1\Sigma_g^+$) and the excited state B ($^3\Pi_{0_u^+}$). Narrow lines are observed in the spectral range 520-498 nm, near the dissociation limit of the molecule [2]. They are of particular interest as references in compact sources realized with frequency doubled diode-pumped solid-state lasers based on $Yb^{3+}$-doped materials which cover this spectral region [3, 4].

The width of the hyperfine levels in the excited state $^3\Pi_{0_u^+}$ is limited by their radiative width and by the predissociation effects resulting from the gyroscopic and hyperfine couplings with the dissociative state $^1\Pi_{1_u}$ [5, 6]. The natural predissociation becomes negligible for the high vibrational levels [7] and the width of the hyperfine levels are then essentially given by their radiative width, which decreases near the dissociation limit of the molecule [5]. Lifetime measurements of rovibronic states near this limit by fluorescence decay have shown that hyperfine transitions with a natural width lower than 10 kHz are expected near 500 nm. In a previous work, we have obtained a half-width at half-maximum (HWHM) experimental linewidth of about 30 kHz for the R(26) 62-0 transition, accessible with an Argon laser at 501.7 nm [8] and recently we used such a narrow line detected in a cell for the long term stabilisation of the laser [9].

The hyperfine splittings of the R(26) 62-0 transition was first measured by Sorem *et al.* in the 70's with an uncertainty of about 1 MHz [10]. More recently, a detailed study of the iodine hyperfine structure at 514.5 nm and 501.7 nm was reported in [11]. The frequency differences between the components of the R(26) 62-0 transition were measured with an uncertainty of 5 kHz.

In the present work, we propose a more accurate determination of the hyperfine splittings of the R(26) 62-0 transition. It is based on the direct frequency measurement of an Ar$^+$ laser locked successively to every hyperfine components of the transition. They are detected in a continuously pumped cell by saturated absorption spectroscopy at low pressure (0.33 Pa). Thus, narrow resonances with a HWHM of about 50 kHz are used as frequency references. The uncertainty of the frequency measurements is lower than 1 kHz.

**Experimental set-up**

Some modifications were made to our Ar$^+$/I$_2$ standard compared with the set-up used for the first absolute frequency measurement of the component a$_7$ of the R(26) 62-0 transition at 501.7 nm [9] whereas the system for absolute frequency measurement is essentially the same. We only give here a brief description of the set-up (Fig. 1) and we specify the differences with the previous system.

The single mode Ar$^+$ laser is first prestabilized to a Fabry-Perot resonator mode by the Pound-Drever-Hall technique, in order to reduce its frequency jitter. In [9], it was then locked to a hyperfine component of the R(26) 62-0 transition detected in a low-pressure iodine cell for long term stabilisation. In fact, it was observed that the stability was limited by mechanical vibrations of the Fabry-Perot cavity at a few hertz. To overcome this effect, the laser is now locked to the iodine transition detected in saturated absorption spectroscopy in another cell at a higher pressure (3.3 Pa). The error signal is obtained by the modulation transfer technique: the pump beam is frequency modulated at 125 kHz and the first harmonic of the saturated absorption signal is detected on the probe beam. In order to improve the signal-to-noise ratio (SNR), the probe beam power is stabilized in a narrow frequency band centred at 125 kHz [12, 9].

The Ar$^+$ laser, prestabilized to the Fabry-Perot resonator mode and the high-pressure iodine cell is then locked to the same hyperfine component of the R(26) 62-0 transition detected in the low-pressure iodine cell. The resonance is detected in saturated absorption spectroscopy with the technique of frequency modulation (FM) spectroscopy [13, 14]. The probe beam is frequency-modulated at $f_1 = 2.5$ MHz with an acousto-optic modulator (AOM). The residual amplitude modulation generated by the AOM is cancelled by the narrow-band control of the beam intensity at the modulation frequency [12]. The phase sensitive detection of the probe beam at 2.5 MHz gives an error signal with a dispersion shape. An additional amplitude modulation of the saturation beam at $f_2 = 200$ Hz and a second phase sensitive detection are used to eliminate the residual offset of the error signal due to the Doppler background.

The length of the low pressure cell is 4 m, its diameter is 15 cm. The beam (diameter 6 mm) passes two times through it to reach an 8 m interaction length with the iodine vapour. This cell is pumped during the experiment to minimize buffer gases and impurities effects. Indeed, the width of the studied transitions at 501.7 nm is known to be very sensitive to the impurities present in the cell [15]. The iodine pressure is controlled by thermostabilization of a cold finger. The narrow lines detected with this cell in previous works using Raman and Rayleigh spectroscopy [16, 17] show that the upper limit of impurity pressure in the cell is less than 0.04 Pa (0.3 mTorr).

The frequency of the $Ar^+$ laser locked to the studied hyperfine component is compared to a mode of a wide bandwidth optical comb generated by a Kerr-lens mode-locked Ti:Sa laser spectrally enlarged by a photonic crystal fibre. One detects the beat note signal between a small amount of the laser power (about 10 mW) and the optical comb. It is then electronically shifted by the carrier envelop offset frequency, detected with the usual f-2f set-up, and is measured with a frequency counter (for details, see [9]). A local oscillator locked to a 1 GHz reference drives the frequency counter and the repetition rate of the femtosecond laser. The reference signal with a high level of stability and accuracy (cryogenic oscillator/H-maser/Cs-fountain) [18] is received from LNE-SYRTE at Observatoire de Paris through a 43 km long optical fibre, as an amplitude modulation of a 1.5 μm optical carrier. The whole frequency measurement set-up has a resolution of a few $10^{-14}$ for 1 s integration time [19].

## Measurement of the hyperfine structure of the R(26)62-0 transition

The frequency standard at 501.7 nm shows a stability better than $10^{-12}$ for 1 s integration time, limited by the signal to noise ratio of the detected line in the low pressure cell. Several factors can affect the frequency reproducibility. The frequency shifts with the iodine pressure, the beam power and the modulation index were investigated in [9].

- The iodine pressure in the cell is controlled by the temperature of a cold finger. The temperature is -22.5°C corresponding to an iodine pressure of 0.33 Pa. The HWHM linewidth of the resonance is then 45 kHz (Fig. 2). The pressure shift coefficient is less than $-3.8\,\text{kHz/Pa}$. At -22.5°C, the sensitivity of the iodine pressure with temperature is 0.04 Pa/K, which gives a very low sensitivity of the measured frequency with the temperature of about -150 Hz/K.

- The probe beam power is 0.5 mW. Its decrease by a factor 2 leads to a shift of about 1 kHz. That corresponds to a sensitivity of the frequency with the power of 4 kHz/mW.

- The modulation index is close to unity which is the optimum value for the signal detection. The modification of the modulation index does not lead to any appreciable shift.

Considering an uncertainty of the cold finger temperature of 1 K, the frequency uncertainty associated to the iodine pressure is 150 Hz. The uncertainty of the beam power is estimated to be 10 %, corresponding to a frequency uncertainty of 200 Hz. The frequency measurement chain does not introduce any frequency error. The combined standard uncertainty (1σ) is then $\sqrt{150^2 + 200^2} = 250\,\text{Hz}$.

In a first step, several measurements of the absolute frequency of component $a_7$ of the R(26) 62-0 transition were repeated during two successive days. It is seen in Fig. 3 that the deviations between measurements operated from one day to the other does not exceed the deviations observed during measurements operated on the same day ($\leq 250\,\text{Hz}$). The mean value of the 8 absolute frequency measurements of component $a_7$ is found to be 597 366 498 654.47 kHz.

The standard deviation is 160 Hz, a smaller value than the combined standard uncertainty of 250 Hz estimated above.

In a second step, the laser was locked successively on the fifteen hyperfine components of the R(26) 62-0 transition and their absolute frequencies were measured. The absolute frequency of component $a_1$ is found to be 597 366 101 477.9 kHz. The positions of the other lines, referred to component $a_1$ are reported in Table 1.

The hyperfine splittings of the R(26) 62-0 transition deduced from our absolute frequency measurements can be compared to data reported in [11]. In this work, the frequency differences between hyperfine components were obtained from the differences of the beat frequency of two $Ar^+$ lasers, one of them locked to a reference component, and the other one locked successively to the hyperfine components of the transition. We calculated the difference between the splittings deduced in the present work from absolute frequency measurements and those given in [11]. The global frequency shift between both sets of data is compensated by taking the sum of the deviations equal to 0. The resulting values are given in the last column of Table 1. The maximum deviation between both sets of data is 12.2 kHz and the standard deviation is 6.8 kHz. This latter value is of the same order as the uncertainty of the frequency differences (5 kHz - 1 $\sigma$) estimated in [11] from the statistical reproducibility.

**Conclusion**

We have performed for the first time, the absolute frequency measurement of the complete hyperfine structure of the R(26) 62-0 transition in molecular iodine at 501.7 nm with an optical clockwork based on a femtosecond laser frequency comb generator. Our $Ar^+$ laser was locked to narrow resonances of about 50 kHz linewidth, detected in a low pressure continuously pumped cell with a low impurity pressure. The uncertainty of the frequency measurements is $\sigma = 250 \, \text{Hz}$.

These measurements give a better accuracy on the hyperfine splittings of the R(26) 62-0 transition. These data can be combined with the measurement of hyperfine splittings of the lower level of the R(26) 62-0 transition studied with Raman technique [20] in order to determine precisely the hyperfine interaction constants of the upper level v' = 62 near the dissociation limit.

## Acknowledgments

The authors wish to thank Ch. J. Bordé for his interest in this work.

This work was supported by GDRE project "Lasers et techniques de l'information" and by RFBR grant #05-02-19645.

**Table captions**

**Tab. 1:** Absolute frequencies of the hyperfine components of the R(26) 62-0 transition in iodine and deviations between the frequencies deduced from [11] and our frequencies, taking the sum of the deviations equal to 0.

**Figure captions**

**Fig. 1:** $Ar^+/I_2$ frequency standard at 501.7 nm.

AOM: acousto-optic modulator; PD: photodetector; FP: Fabry-Perot resonator; AM: amplitude modulation, FM: frequency modulation.

**Fig. 2:** Error signal for the long term stabilization of the $Ar^+$ laser on the hyperfine component $a_7$ of the R(26) 62-0 transition. Experimental conditions: iodine pressure: 0.33 Pa; saturating beam power: 2.2 mW; probe beam power: 0.5 mW; beams diameter: 6 mm; time constant: 300 ms; one sweep of 60 s.

**Fig. 3:** Frequency measurements of the hyperfine component $a_7$ of the R(26) 62-0 transition. The mean value is $f_{mean}$ = 597 366 498 654.47 kHz.

Table 1

| component | $f(a_n)$ – 597 366 101 477.9 kHz (kHz) | Frequencies deduced from [11] - ours (kHz) |
|---|---|---|
| $a_1$ | 0 | 6.5 |
| $a_2$ | 166 253.5 | 5.0 |
| $a_3$ | 265 934.0 | 1.6 |
| $a_4$ | 295 233.9 | -4.4 |
| $a_5$ | 317 938.2 | -1.6 |
| $a_6$ | 364 499.5 | -11.0 |
| $a_7$ | 397 176.6 | 11.0 |
| $a_8$ | 474 039.1 | -0.6 |
| $a_9$ | 523 548.2 | -6.7 |
| $a_{10}$ | 560 031.4 | -2.9 |
| $a_{11}$ | 637 054.1 | -5.6 |
| $a_{12}$ | 660 262.7 | 3,8 |
| $a_{13}$ | 739 965.5 | 1.0 |
| $a_{14}$ | 769 747.2 | -8.7 |
| $a_{15}$ | 842 244.3 | 12.2 |

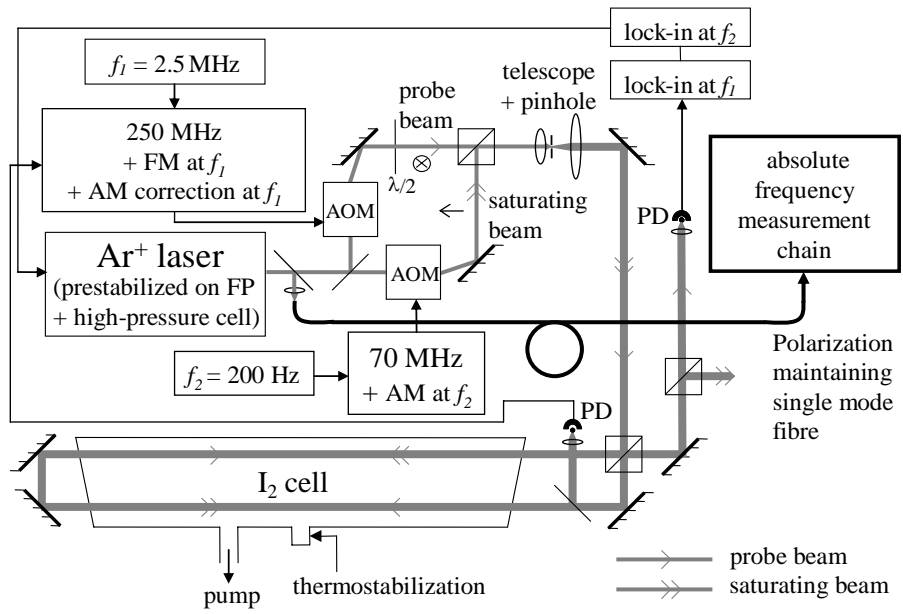

Fig. 1

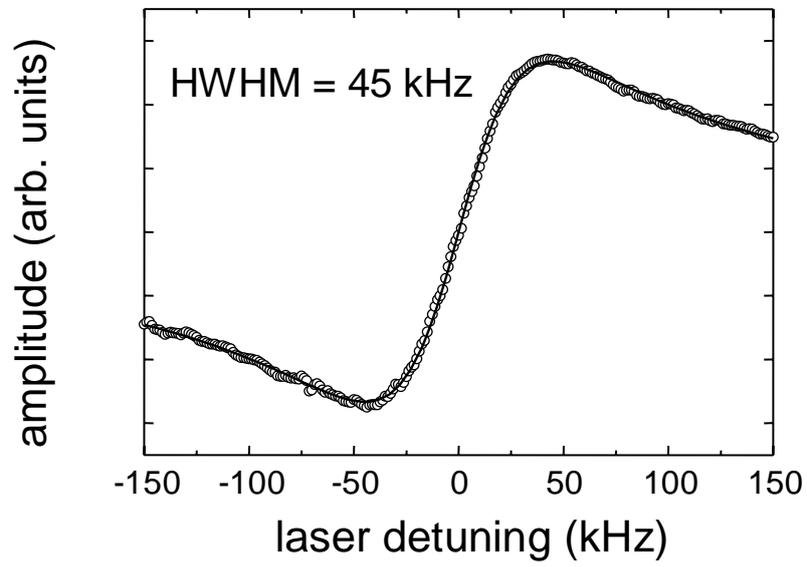

Fig. 2

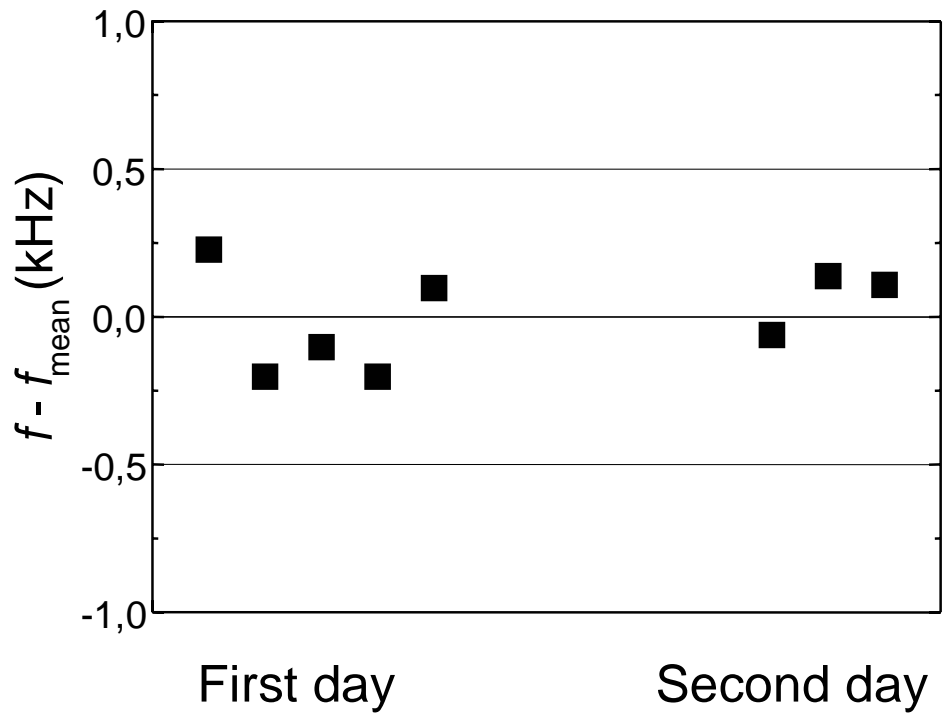

Fig. 3